\documentclass[english,aps,prl,reprint]{revtex4-1}
\usepackage[T1]{fontenc}
\usepackage[latin9]{inputenc}
\usepackage{bm}
\usepackage{amsmath}
\usepackage{graphicx}

\makeatletter
\usepackage[all]{xy}

\usepackage{variables}
\usepackage[bookmarks=false]{hyperref}

\makeatother

\usepackage{babel}
\begin{document}
\title{Mott transition from the non-analyticity of the one-body reduced density-matrix
functional}
\author{Zhengqian Cheng and Chris A. Marianetti}
\affiliation{Department of Applied Physics and Applied Mathematics, Columbia University,
New York, NY 10027}
\date{\today}
\begin{abstract}
One-body reduced density-matrix functional (1RDMF) theory has yielded
promising results for small systems such as molecules, but has not
addressed quantum phase transitions such as the Mott transition. Here
we explicitly execute the constrained search within a variational
ansatz to construct a 1RDMF for the multi-orbital Hubbard model with
up to seven orbitals in the thermodynamic limit. The variational ansatz
is the $\mathcal{N}=3$ ansatz of the variational discrete action
theory (VDAT), which can be exactly evaluated in $d=\infty$. The
resulting 1RDMF exactly encapsulates the $\mathcal{N}=3$ VDAT results,
which accurately captures Mott and Hund physics. We find that non-analytic
behavior emerges in our 1RDMF at fixed integer filling, which gives
rise to the Mott transition. We explain this behavior by separating
the constrained search into multiple stages, illustrating how a nonzero
Hund exchange drives the continuous Mott transition to become first-order.
Our approach creates a new path forward for constructing an accurate
1RDMF for strongly correlated electron materials.
\end{abstract}
\maketitle
\global\long\def\rhoeff{\bm{\rho}}%

One-body reduced density matrix functional (1RDMF) theory can be viewed
as a formalism to encapsulate the ground state properties for a class
of Hamiltonians with fixed interactions and arbitrary one-body terms
\cite{Gilbert19752111,Donnelly19784431,Levy19796062,Piris2007134,Pernal2016125}.
There have been successes in developing explicit interaction energy
functionals which produce reasonable energetics in various molecules
\cite{Kutzelnigg19643640,Cioslowski20036443,Piris2011164102,Pernal2013127,Piris20141169,Pernal2016125,Schade20172677}
and single orbital Hubbard models \cite{Lopezsandoval20001764,Lopezsandoval2002155118,Lopezsandoval2004085101,Saubanere2011035111,Tows20131422,Kamil2016085141,Saubanere2016045102,Mitxelena2017425602,Mitxelena2018089501,Mitxelena2020064108,Mitxelena20201701},
in addition to select observables in crystals \cite{Helbig200767003,Sharma2008201103,Lathiotakis2009040501,Sharma2013116403,Shinohara2015093038}.
However, there has not been a rigorous demonstration that any existing
1RDMF can capture the Mott transition. Our focus is capturing local
Mott and Hund physics, which is hosted in strongly correlated electron
materials (SCEM) which often bear $d$ or $f$ electrons \cite{Kent2018348}.
The essence of SCEM is embodied by the multi-orbital Hubbard model
in infinite dimensions, which facilitates numerically exact solutions
and serves as a reasonable starting approximation for crystals in
two and three dimensions \cite{Georges199613,Kotliar200453,Kotliar2006865,Kent2018348}.
A robust 1RDMF of the multi-orbital Hubbard model should satisfy the
following criteria, which are dictated by numerically exact solutions
in infinite dimensions. First, the Mott transition must be predicted
at a finite Hubbard $U$. Second, the order of the Mott transition
must be properly predicted with or without the Hund coupling $J$.
Third, the aforementioned goals should be achieved while maintaining
the symmetry of the Hamiltonian, proving that the functional can faithfully
describe Mott and Hund physics. Here we demonstrate that an accurate
1RDMF that satisfies the preceding criteria can be constructed by
\textit{exactly} executing the constrained search within the $\mathcal{N}=3$
sequential product density matrix (SPD) ansatz of the variational
discrete action theory (VDAT) in infinite dimensions \cite{Cheng2021206402,Cheng2021195138}.
Given that VDAT at $\mathcal{N}=3$ has been demonstrated to accurately
capture Mott and Hund physics \cite{Cheng2022205129,Cheng2023035127},
the resulting 1RDMF serves as an efficient approach to exactly encapsulate
VDAT results. A companion article to this Letter introduces the qubit
parameterization of VDAT, which provides the technical tools needed
to execute the constrained search \cite{companion}.

For simplicity, we consider the translationally invariant Hamiltonian
$\hat{H}=\sum_{k\ell}\epsilon_{k\ell}\hat{n}_{k\ell}+\hat{H}_{int}$,
where $k$ labels a reciprocal lattice point, $\ell$ enumerates $2N_{orb}$
spin orbitals, and the interaction Hamiltonian is $\hat{H}_{int}=\sum_{i}H_{loc}\left(\left\{ \hat{n}_{i\ell}\right\} \right)$
where $H_{loc}$ is an arbitrary polynomial function and $\hat{n}_{i\ell}$
is the density operator for spin orbital $\ell$ at site $i$. It
should be noted that the one-body contribution is restricted to be
diagonal in $\ell$, which is a common simplification used when studying
the multi-orbital Hubbard model (e.g. see Ref. \cite{Chatzieleftheriou2023066401}).
For this class of Hamiltonians, the interaction energy functional
is defined as 
\begin{align}
 & E_{int}\left(\{n_{k\ell}\}\right)=\frac{1}{L}\min_{|\Psi\rangle}\left\{ \langle\Psi|\hat{H}_{int}|\Psi\rangle\big|\langle\Psi|\hat{n}_{k\ell}|\Psi\rangle=n_{k\ell}\right\} ,\label{eq:constrained_search}
\end{align}
where $|\Psi\rangle$ is a normalized state in the many-particle Fock
space, $L$ is the number of sites in the lattice, and $n_{k\ell}\in[0,1]$.
For a given $\hat{H}_{int}$, the interaction functional $E_{int}$
is universal in the sense that it provides the ground state energy
for Hamiltonians with arbitrary $\epsilon_{k\ell}$ as follows 
\begin{equation}
E=\min_{\{n_{k\ell}\}}(\frac{1}{L}\sum_{k\ell}\epsilon_{k\ell}n_{k\ell}+E_{int}\left(\{n_{k\ell}\}\right)).
\end{equation}
While $E_{int}\left(\{n_{k\ell}\}\right)$ must be explicitly constructed
for a given $\hat{H}_{int}$, the interacting energy functional for
$\lambda\hat{H}_{int}$, where $\lambda$ is a positive number, is
the interacting energy functional for $\hat{H}_{int}$ multiplied
by $\lambda$. 

\begin{figure}
\includegraphics[width=0.99\columnwidth]{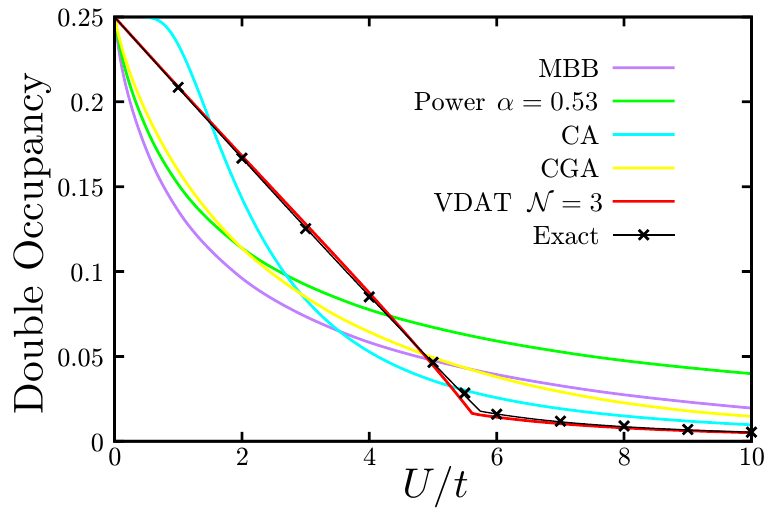}\caption{\label{fig:The-double-occupancy-single-band}The double occupancy
vs. $U/t$ for the single-orbital Hubbard model on the Bethe lattice
in $d=\infty$ using four 1RDMF's: MBB, Power, CA, and CGA. The exact
(DMFT) and VDAT results are taken from Ref. \cite{Cheng2021206402}.
The black line is obtained from interpolations of the metallic and
insulating DMFT data. }
\end{figure}

We begin by examining the fidelity of numerous explicit 1RDMF's from
the literature by applying them to the single-orbital Hubbard model
in $d=\infty$, including the MBB \cite{Muller1984446}, CA \cite{Csanyi20007348},
CGA \cite{Csanyi2002032510}, and power \cite{Sharma2008201103} functionals,
all of which approximate the interaction energy functional as a Hartree
term plus an exchange-correlation term which consists of independent
contributions from each spin orbital. We compare all functionals to
numerically exact solutions obtained using dynamical mean-field theory
(DMFT) \cite{Georges199613,Kotliar200453} with the numerical renormalization
group (NRG) impurity solver \cite{Zitko2009085106}, in addition to
VDAT results for $\mathcal{N}=3$ \cite{Cheng2021206402} (see Figure
\ref{fig:The-double-occupancy-single-band}). The VDAT results correctly
describe the continuous Mott transition \cite{Georges199613}, identified
by a kink in the double occupancy at $U/t=5.61784$, and accurately
agree with the DMFT results. The MBB, CA, CGA, and power functionals
are all substantially in error, and qualitatively fail to predict
the Mott transition at a finite $U/t$. More advanced functionals,
including PNOF5 \cite{Piris2011164102} and PNOF7 \cite{Piris2017063002},
also perform poorly in the absence of symmetry breaking \cite{companion}.

Our proposed solution for obtaining the interaction energy functional
is to rigorously execute the constrained search within the sequential
product density matrix (SPD) ansatz \cite{Cheng2021206402,Cheng2021195138},
which is the ansatz used in VDAT. The variational freedom of the SPD
is set by an integer $\mathcal{N}$, where $\mathcal{N}\rightarrow\infty$
encapsulates the exact solution, and there are two types of SPD: G-type
or B-type. A restricted form of the SPD has been evaluated using quantum
Monte-Carlo to solve the single-orbital Hubbard model in two dimensions
\cite{Otsuka19921645,Yanagisawa19983867,Yanagisawa2016114707,Yanagisawa2019054702,Sorella2023115133,Levy2024013237},
the $p$-$d$ model \cite{Yanagisawa202127004,Yanagisawa2001184509},
and selected molecules \cite{Chen20234484}. Alternatively, VDAT has
been used to exactly evaluate the SPD in the Anderson impurity model
\cite{Cheng2021206402}, the single orbital Hubbard model in $d=\infty$
\cite{Cheng2021206402}, and the multi-orbital Hubbard model in $d=\infty$
\cite{Cheng2022205129,Cheng2023035127}. Nonetheless, constructing
$E_{int}\left(\{n_{k\ell}\}\right)$ from any non-trivial variational
ansatz is still highly challenging as there must be some efficient
means to constrain the variational parameters to a particular $\{n_{k\ell}\}$
and minimize over the remaining degrees of freedom. The recently developed
qubit parameterization of VDAT allows for the constrained search to
be executed for the B-type $\mathcal{N}=2$ and the G-type $\mathcal{N}=2$
and $\mathcal{N}=3$ SPD, and here we focus on G-type $\mathcal{N}=3$
(see Ref. \cite{companion} for $\mathcal{N}=2$).

We now proceed to construct the interaction energy functional via
the $\mathcal{N}=3$ G-type SPD, and it is sufficient to restrict
the SPD to a pure state which can be represented as
\begin{equation}
|\Psi\rangle=\exp(\sum_{k\ell}\gamma_{k\ell}\hat{n}_{k\ell})\exp(\sum_{i\Gamma}\upsilon_{\Gamma}\hat{X}_{i\Gamma})|\Psi_{0}\rangle,\label{eq:GBWF}
\end{equation}
where $|\Psi_{0}\rangle$ is a Slater determinant, $\hat{X}_{i\Gamma}$
is a diagonal Hubbard operator at site $i$, and the variational parameters
are $\{\gamma_{k\ell}\}$, $\{\upsilon_{\Gamma}\}$, and $\{n_{k\ell;0}\}$,
where $n_{k\ell;0}=\langle\Psi_{0}|\hat{n}_{k\ell}|\Psi_{0}\rangle\in\{0,1\}$.
The index $\Gamma$ enumerates all $2^{2N_{orb}}$ local atomic states.
The key idea is to reparametrize the variational parameters $\{\gamma_{k\ell}\}$
and $\{\upsilon_{\Gamma}\}$ from Eq. \ref{eq:GBWF} in terms of new
variational parameters $\{n_{k\ell}\}$ and $\rhoeff$. The $\{n_{k\ell}\}$
is the physical momentum density distribution taking values within
$[0,1]$, and has the same number of free parameters as $\{\gamma_{k\ell}\}$.
The $\rhoeff$ is a many-body density matrix corresponding to a pure
state of a $2N_{orb}$ qubit system, which has the same number of
independent parameters as $\{\upsilon_{\Gamma}\}$. The qubit system
is a convenient mathematical tool to represent local integer time
correlation functions of the VDAT formalism as static observables
\cite{companion}. There is a natural correspondence between the density
operators in the local Fock space and the qubit space, where the local
density operator is represented as $\hat{n}_{\ell}=\frac{1}{2}\left(1-\hat{\sigma}_{\ell}^{z}\right)$,
and $\hat{\sigma}_{\ell}^{\mu}$ denotes the application of the $\hat{\sigma}^{\mu}$
Pauli matrix to the $\ell$-th qubit subspace. Using these new variables,
the interaction energy for a given set of variational parameters can
be expressed as $\langle\hat{H}_{eff}\rangle_{\rhoeff}$, where $\hat{H}_{eff}=H_{loc}\left(\left\{ \hat{n}_{eff,\ell}\right\} \right)$
is defined in the qubit space, and $\hat{n}_{eff,\ell}$ depends on
the variational parameters only through the following five variables:
$n_{\ell}=\int dkn_{k\ell}$, $\xi_{\ell}=\frac{1}{2}\langle\hat{\sigma}_{\ell}^{x}\rangle_{\rhoeff},$
$\Delta_{\ell}=\int_{>}dkn_{k\ell}$, and $\mathcal{A}_{X\ell}=\int_{X}dk\sqrt{n_{k\ell}\left(1-n_{k\ell}\right)}$,
where $X$ takes two values denoted as $<$ or $>$ indicating that
the integration is over the region where $n_{k\ell,0}=1$ or $n_{k\ell,0}=0$
for a given $\ell$, respectively. Here we have taken the continuum
limit of the discretized $n_{k\ell}$ and choose the convention $\int dk=1$.
It should be emphasized that $\hat{n}_{eff,\ell}$ \textit{analytically}
depends on the preceding five variables (see Eq. 68 in Ref. \cite{companion}).
Finally, the interaction energy functional can be conveniently expressed
as a constrained minimization

\begin{align}
 & E_{int}\left(\{n_{k\ell}\}\right)=\min_{\rhoeff,\{n_{k\ell;0}\}}\langle\hat{H}_{eff}\rangle_{\rhoeff},\label{eq:qubit_interaction_energy}\\
 & \textrm{subject to}\hspace{1em}\int dkn_{k\ell;0}=\left\langle \frac{1-\hat{\sigma}_{\ell}^{z}}{2}\right\rangle _{\rhoeff}=\int dkn_{k\ell},\label{eq:constraint1}\\
 & \left|\langle\hat{\sigma}_{\ell}^{x}\rangle_{\rhoeff}\right|\leq2\sqrt{\left(1-n_{\ell}\right)n_{\ell}-\left(1-\Delta_{\ell}\right)\Delta_{\ell}},\label{eq:constraint2}
\end{align}
which corresponds to an \textit{exact} constrained search within the
wave function given by Eq. \ref{eq:GBWF} in $d=\infty$. We assume
that the optimized $n_{k\ell;0}$ is given as $\theta(n_{k\ell}-n_{\ell}^{\star})$,
where $\theta$ is the Heaviside function and $n_{\ell}^{\star}$
is chosen such that $\int dkn_{k\ell;0}=n_{\ell}$. The remaining
challenge is how to efficiently minimize $\langle\hat{H}_{eff}\rangle_{\rhoeff}$
over $\rhoeff$, which is non-trivial given that $\hat{H}_{eff}$
depends on $\rhoeff$ through $\xi_{\ell}$. This challenge can be
addressed by dividing the minimization into two stages. First, $\{\xi_{\ell}\}$
can be fixed, which will fix $\hat{H}_{eff}$, and then $\langle\hat{H}_{eff}\rangle_{\rhoeff}$
can be minimized, which is equivalent to finding the ground state
of $\hat{H}_{eff}-\sum_{\ell}h_{\ell}^{x}\hat{\sigma}_{\ell}^{x}-\sum_{\ell}h_{\ell}^{z}\hat{\sigma}_{\ell}^{z}$,
where the Lagrange multipliers $h_{\ell}^{\mu}$ are chosen to yield
$\langle\hat{\sigma}_{\ell}^{x}\rangle_{\rhoeff}=2\xi_{\ell}$ and
$\langle\hat{\sigma}_{\ell}^{z}\rangle_{\rhoeff}=1-2n_{\ell}$. Second,
the interaction energy functional is obtained by minimizing over $\{\xi_{\ell}\}$.
In what follows, we will illustrate this two stage minimization for
the case where $n_{k\ell}$ is restricted to be independent of $\ell$
and have particle-hole symmetry, which is sufficient to solve a multi-orbital
Hubbard model where $\epsilon_{k\ell}$ is independent of $\ell$
and has particle-hole symmetry. 

We consider the following multi-orbital interaction Hamiltonian $H_{loc}\left(\left\{ \hat{n}_{\ell}\right\} \right)=U\hat{O}$,
where
\begin{equation}
\hat{O}=\hat{O}_{1}+\left(1-2\frac{J}{U}\right)\hat{O}_{2}+\left(1-3\frac{J}{U}\right)\hat{O}_{3},
\end{equation}
where $\hat{O}_{1}=\sum_{\alpha}\delta\hat{n}_{\alpha\uparrow}\delta\hat{n}_{\alpha\downarrow}$,
$\hat{O}_{2}=\sum_{\alpha<\beta,\sigma}\delta\hat{n}_{\alpha\sigma}\delta\hat{n}_{\beta\bar{\sigma}}$,
$\hat{O}_{3}=\sum_{\alpha<\beta,\sigma}\delta\hat{n}_{\alpha\sigma}\delta\hat{n}_{\beta\sigma}$,
and $\delta\hat{n}_{\alpha\sigma}=\hat{n}_{\alpha\sigma}-\frac{1}{2}$,
with the orbital indices $\alpha,\beta$ taking values of $1,\dots,N_{orb}$
and the spin index $\sigma\in\{\uparrow,\downarrow\}$. The restriction
of particle-hole symmetry at half-filling ensures that $\mathcal{A}_{<\ell}=\mathcal{A}_{>\ell}$
and $n_{\ell}=\frac{1}{2}$, and for convenience we define $A=\mathcal{A}_{<\ell}+\mathcal{A}_{>\ell}$
and also discard the index $\ell$ for $\xi_{\ell}$ and $\Delta_{\ell}$.
The interaction energy functional implicitly depends on $N_{orb}$,
$U$, and $J$, but scaling the interaction Hamiltonian will trivially
scale the interaction energy functional. Therefore, the interaction
energy functional only has a non-trivial dependence on $N_{orb}$
and $J/U$. 

\begin{figure}
\includegraphics[width=0.99\columnwidth]{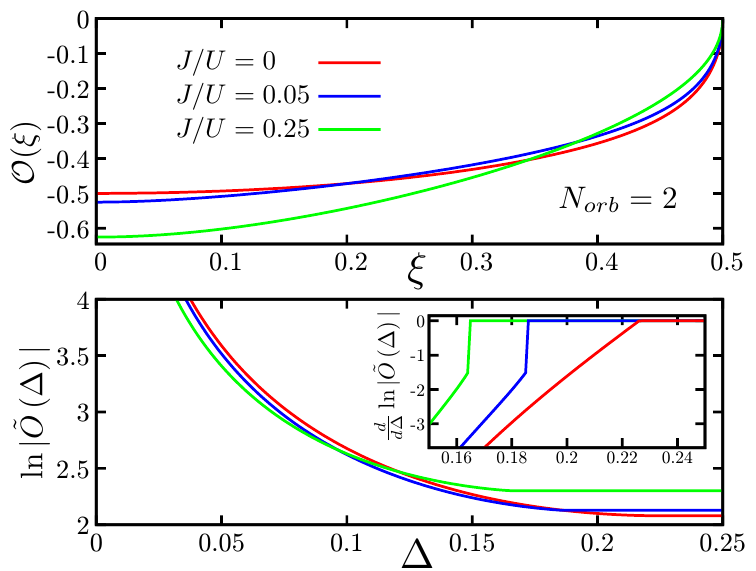}

\caption{\label{fig:fancy_O_xi} Plots of $\mathcal{O}(\xi)$ (top) and $\ln|\tilde{O}\left(\Delta\right)|$
(bottom) for $N_{\textrm{orb}}=2$ and various $J/U$; the inset plots
$\frac{d}{d\Delta}\ln|\tilde{O}\left(\Delta\right)|$, highlighting
the non-analytic behavior in the 1RDMF. }
\end{figure}

The interaction energy per site can be written as $\langle\hat{H}_{eff}\rangle_{\rhoeff}=UA^{4}\mathcal{\tilde{F}}^{2}(\Delta,\xi)\langle\hat{O}\rangle_{\rhoeff}$
(see Eq. 431 in Ref. \cite{companion}), where
\begin{align}
 & \mathcal{\tilde{F}}(\Delta,\xi)=\frac{2}{1-4\xi^{2}}\left(\sqrt{1-\frac{4\xi^{2}}{\left(1-2\Delta\right){}^{2}}}+1\right).\label{eq:fancy-F}
\end{align}
Following the two stage minimization outlined above, we first minimize
the interaction energy over $\rhoeff$ under a constrained value of
$\xi$ by computing 
\begin{align}
 & \mathcal{O}(\xi)=\min_{\rhoeff}\{\langle\hat{O}\rangle_{\rhoeff}|\frac{1}{2}\langle\hat{\sigma}_{\ell}^{x}\rangle_{\rhoeff}=\xi,\hspace{0.2em}\langle\hat{\sigma}_{\ell}^{z}\rangle_{\rhoeff}=0\},\label{eq:fancyO_xi}
\end{align}
which encodes the essential information about all local correlations.
The computational cost for evaluating $\mathcal{O}(\xi)$ will generally
scale exponentially in terms of $N_{orb}$, but it will be demonstrated
that only a coarse disctretization in $\xi$ is needed to faithfully
represent $\mathcal{O}(\xi)$. A given data point can be generated
efficiently by solving the ground-state of $\hat{O}-h\sum_{\ell}\hat{\sigma}_{\ell}^{x}$,
where $h$ is a Lagrange multiplier, which is a transverse field Ising
model. In Figure \ref{fig:fancy_O_xi}, $\mathcal{O}(\xi)$ is plotted
for $N_{orb}=2$ and various $J/U$, demonstrating that the result
is indeed a smooth function which can be accurately represented by
a spline interpolation. Given that $\mathcal{O}(\xi)$ is demonstrated
to be an analytic function when $|\xi|<\frac{1}{2}$, any non-analytic
behavior in the interaction energy must arise from minimizing over
$\xi$. The interaction energy functional can be \textit{explicitly}
constructed when $n_{k\ell}$ is independent of $\ell$ and has particle
hole symmetry, given as
\begin{align}
 & E_{int}\left(\{n_{k\ell}\}\right)=UA^{4}\tilde{O}\left(\Delta\right),\label{eq:Eint_half_fill}\\
 & \tilde{O}\left(\Delta\right)=\min_{\xi\in\left[0,\frac{1}{2}-\Delta\right]}\tilde{\mathcal{F}}^{2}\left(\Delta,\xi\right)\mathcal{O}(\xi),\label{eq:Odelta}
\end{align}
where $\tilde{O}\left(\Delta\right)$ can be efficiently evaluated
using the spline of $\mathcal{O}(\xi)$, and the result can be accurately
represented with a spline interpolation as well. Moreover, for the
case of $N_{orb}=1$, an analytical form of $\tilde{O}\left(\Delta\right)$
can be derived (see Eq. (103) in Ref. \cite{companion}). Given that
$\tilde{O}\left(\Delta\right)$ is constructed for a given $J/U$
and $N_{orb}$, the result may be used to solve any model with $\epsilon_{k\ell}$
independent of $\ell$ with particle-hole symmetry and $U$. In Figure
\ref{fig:fancy_O_xi}, we plot $\ln|\tilde{O}\left(\Delta\right)|$
and $d\ln|\tilde{O}\left(\Delta\right)|/d\Delta$ for $N_{orb}=2$
and various $J/U$, demonstrating that there is a discontinuity at
$\Delta=\Delta_{c}$ in the first derivative for $J/U>0$ and the
second derivative for $J/U=0$. The presence of non-analyticity in
the interaction energy functional is profound, as it guarantees that
there must be a phase transition for an arbitrary band structure at
some critical $U$, and this transition will be demonstrated to be
the Mott transition. This non-analytic behavior in the interaction
energy functional should not be confused with the derivative discontinuity
in density functional theory \cite{Perdew19821691}, which is associated
with a change in the total number of particles and is present for
all systems with a charge gap. 

\begin{figure}
\includegraphics[width=0.99\columnwidth]{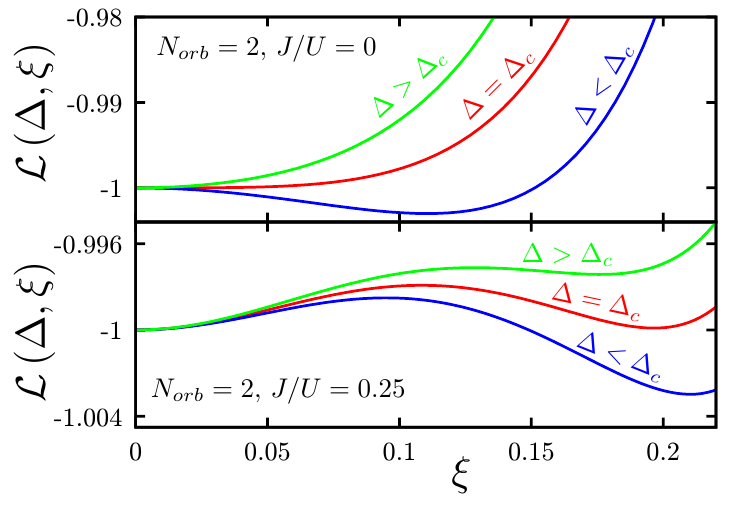}

\caption{\label{fig:L_vs_xi_plots} Graphical illustration of the origin of
the non-analyticity in the interacting energy functional. Plots of
$\mathcal{L}\left(\Delta,\xi\right)$ versus $\xi$ at $N_{orb}=2$
with $J/U=0$ (top) and $J/U=0.25$ (bottom) for various $\Delta$,
where $\Delta_{c}$ corresponds to the non-analytical point in $\tilde{O}\left(\Delta\right)$.}
\end{figure}

We now explore why $E_{int}\left(\{n_{k\ell}\}\right)$ has non-analytic
regions by investigating the minimization over $\xi$. Consider the
function $\mathcal{L}\left(\Delta,\xi\right)=-\frac{\tilde{\mathcal{F}}^{2}\left(\Delta,\xi\right)\mathcal{O}(\xi)}{\tilde{\mathcal{F}}^{2}\left(\Delta,0\right)\mathcal{O}(0)}$,
where $\mathcal{L}\left(\Delta,0\right)=-1$. Finding the minimum
of $\mathcal{L}\left(\Delta,\xi\right)$ will yield the minimum of
$\tilde{\mathcal{F}}^{2}\left(\Delta,\xi\right)\mathcal{O}(\xi)$
used in Eq. \ref{eq:Odelta}, given that $\tilde{\mathcal{F}}\left(\Delta,0\right)=4$
and $\mathcal{O}\left(0\right)=-\frac{1}{4}N_{\text{orb}}\left(1+\left(N_{\text{orb}}-1\right)J/U\right)$.
We first consider $J/U=0$ at $N_{orb}=2$ for various $\Delta$ (see
Figure \ref{fig:L_vs_xi_plots}), demonstrating that the optimized
$\xi$ will continuously go to zero as $\Delta$ increases through
$\Delta_{c}$. Alternatively, for $J/U>0$, the optimized $\xi$ will
discontinuously go to zero as $\Delta$ increases through $\Delta_{c}$.
Therefore, in both cases, $\Delta>\Delta_{c}$ indicates $\xi=0$
and $\frac{d\widetilde{O}\left(\Delta\right)}{d\Delta}=0$ (see Eq.
\ref{eq:fancy-F}), which can be seen in Figure \ref{fig:fancy_O_xi}.
A sixth order Taylor series expansion of $\mathcal{L}\left(\Delta,\xi\right)$
in terms of $\xi$ can be used to provide an analytical understanding
of the non-analyticity in $\tilde{O}\left(\Delta\right)$ (see Section
VII C in Ref. \cite{companion}). 

We now demonstrate that $\frac{d\widetilde{O}\left(\Delta\right)}{d\Delta}=0$
indicates that the system is in the Mott phase. Given that the interaction
energy only depends on $\{n_{k\ell}\}$ via $\Delta$ and $A$, the
total energy can be partially optimized for a fixed $\Delta$ and
$A$, which is achieved by the following momentum density distribution
\cite{Cheng2023035127,companion}
\begin{equation}
n_{k\ell}=\frac{1}{2}\left(1-\frac{\textrm{\textrm{sgn(\ensuremath{\epsilon_{k\ell}})}}a+\epsilon_{k\ell}}{\sqrt{(\textrm{\textrm{sgn(\ensuremath{\epsilon_{k\ell}})}}a+\epsilon_{k\ell})^{2}+b^{2}}}\right),\label{eq:n_epsilon_a_b-1}
\end{equation}
where $a$ and $b$ are Lagrange multipliers that are determined by
$\Delta$ and $A$. The partially optimized energy is a function of
$\Delta$ and $A$, and rewriting the saddle point equations yields
$a=\frac{UA^{4}}{4N_{\text{orb}}}\frac{d\widetilde{O}\left(\Delta\right)}{d\Delta}$
and $b=\frac{-2UA^{3}}{N_{\text{orb}}}\widetilde{O}\left(\Delta\right)$
(see Eqns. (483) and (484) in Ref. \cite{companion}). Given that
the quasiparticle weight $Z=a/\sqrt{a^{2}+b^{2}}$, and that $A>0$
for finite $U$, the only scenario where $Z=0$ is when $\frac{d\widetilde{O}\left(\Delta\right)}{d\Delta}=0$.
Determining the order of the phase transition requires a global comparison
of the total energy between the metal and insulating phases, and the
transition only occurs at $\Delta_{c}$ when the transition is continuous
(see Section VII C in Ref. \cite{companion}).

\begin{figure}

\includegraphics[width=0.99\columnwidth]{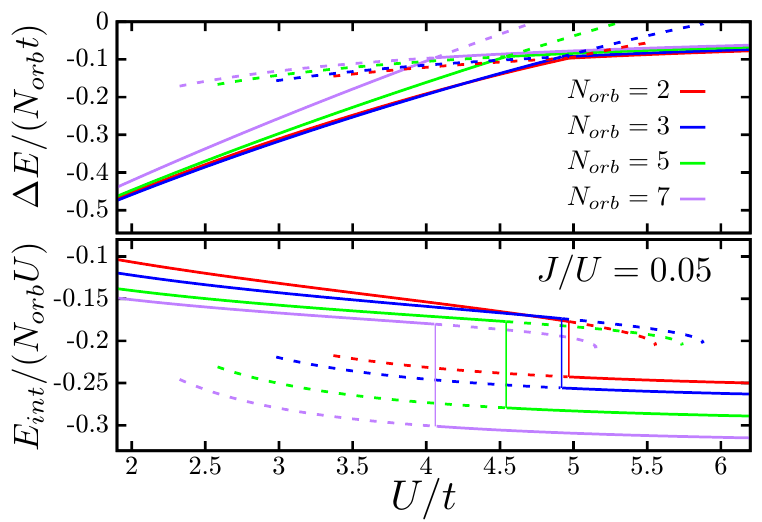}

\caption{\label{fig:multiorbital-energy-vs-U} Plots of various energy contributions
for the multi-orbital Hubbard model on the Bethe lattice in $d=\infty$
for $N_{\textrm{orb}}=2,3,5,7$ as a function of $U/t$ for $J/U=0.05$.
(top) Plot of $\Delta E/(N_{orb}t)$, where $\Delta E=E(U,J,t)-E(U,J,0)$
and $E(U,J,t)$ is the total energy per site. (bottom) Plot of $E_{int}/(N_{orb}U)$.
The dotted lines indicate metastable solutions.}
\end{figure}

\begin{figure}
\includegraphics[width=0.99\columnwidth]{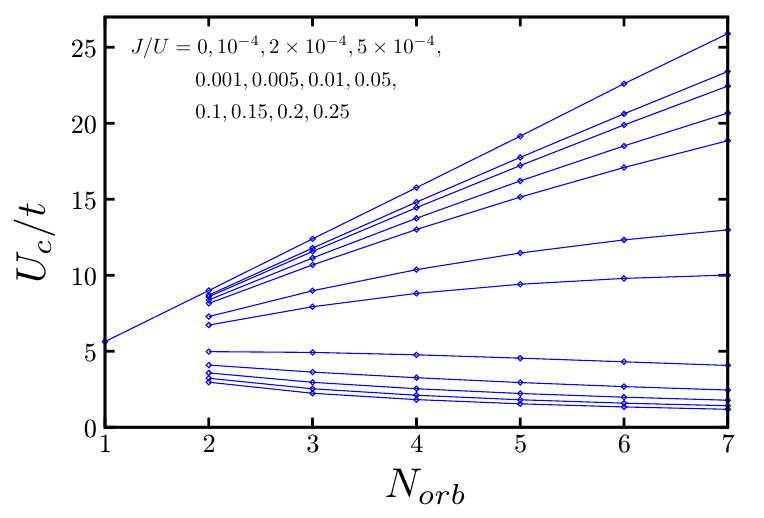}

\caption{\label{fig:Uc} A plot of $U_{c}$ for the multi-orbital Hubbard model
on the Bethe lattice in $d=\infty$ as a function of $N_{\textrm{orb}}$
for various $J/U$, where $U_{c}$ monotonically decreases with increasing
$J/U$ for a given $N_{orb}$.}
\end{figure}

We now proceed to use the 1RDMF to solve the multi-orbital Hubbard
model on the Bethe lattice in $d=\infty$ for $N_{orb}=2-7$. While
these results will be numerically identical to VDAT at $\mathcal{N}=3$
with a G-type SPD, the 1RDMF offers extraordinary practical advantages.
The computational cost of the 1RDMF can be broken into two parts.
First, the interaction energy functional must be constructed for a
given $N_{orb}$ and $J/U$, and the computational cost is dominated
by the solution of a collection of quantum spin models with $2N_{orb}$
spins. Second, the total energy must be minimized for a given Hamiltonian
by solving the saddle point equations for $a$ and $b$, which has
a minimal cost. For a fixed $J/U$ and $N_{orb}$, the interaction
energy functional for arbitrary $U$ can be obtained via scaling,
allowing for the entire parameter space over $U$ to be explored at
a minimal computational cost. The total energy and the interaction
energy are plotted as a function of $U/t$ for $J/U=0.05$ and $N_{orb}=2,3,5,7$
(see Figure \ref{fig:multiorbital-energy-vs-U}). The dotted lines
indicate a metastable metal or insulating phase. The Mott transition
can be identified as a kink in the total energy or a discontinuity
in the interaction energy, which is first-order in this case, consistent
with previous Gutzwiller \cite{Bunemann19974011}, slave boson \cite{Hasegawa19971391},
and DMFT \cite{Ono2003035119} studies. Corresponding results for
$J/U=0,0.25$ are provided in Ref. \cite{companion}. Additionally,
we explore how the Mott transition value $U_{c}$ depends on $N_{orb}$
for many different values of $J/U$ (see Figure \ref{fig:Uc}). For
$J/U=0$, we exactly recover the result from our previous VDAT study
\cite{Cheng2023035127}, which demonstrated excellent agreement with
numerically exact DMFT solutions. For small $J/U$, the $U_{c}$ increases
with $N_{orb}$, while the opposite happens for sufficiently large
$J/U$. While previous DMFT studies already elucidated the fact that
increasing $J/U$ decreases $U_{c}$ at a given $N_{orb}$ for $N_{orb}\le3$
\cite{Han19984199,Ono2003035119,Pruschke2005217,medici2011205112},
our results provide a clear understanding of how $U_{c}$ depends
on $N_{orb}$ for a given $J/U$ up to $N_{orb}=7$. 

In conclusion, we have provided a practical formalism for exactly
executing the constrained search within the G-type $\mathcal{N}=3$
SPD in $d=\infty$ to construct the 1RDMF for the multi-orbital Hubbard
model. We explicitly illustrate that there is non-analyticity in the
interaction energy functional for the multi-orbital Hubbard model
which gives rise to the Mott transition and properly describes the
order of the transition. While the Mott transition in $d=\infty$
has been intensively studied, the 1RDMF demonstrates the universality
of the Mott transition independent of the details of the band structure.
Moreover, the 1RDMF is an efficient tool for accurately solving the
ground state properties of the multi-orbital Hubbard model, yielding
solutions for large $N_{orb}$ which cannot be obtained by competing
methods. 

\textit{Acknowledgements} - This work was supported by the INL Laboratory
Directed Research \& Development (LDRD) Program under DOE Idaho Operations
Office Contract DE-AC07-05ID14517. This research used resources of
the National Energy Research Scientific Computing Center, a DOE Office
of Science User Facility supported by the Office of Science of the
U.S. Department of Energy under Contract No. DE-AC02-05CH11231.

%\bibliographystyle{apsrev4-1}
%\bibliography{refs_from_key_1rdmf,refs_1rdmf}

%

\end{document}